\newcommand{\N}{{\mathbb N}}
\newcommand{\Z}{{\mathbb Z}}
\newcommand{\C}{{\mathbb C}}
\newcommand{\kk}{\mathbf{k}}
\newcommand{\TArrow}[1]
{\ 
\parbox{#1}{\tikz{\draw[{[right]}-{stealth},yshift=2mm,](0,0)--(#1,0);
\draw[{[right]}-{stealth},](0,0)--(#1,0);
}
}
\ }

\def\calA{\mathcal{A}}

\def\calF{\mathcal{F}}
\def\calG{\mathcal{G}}
\def\calJ{\mathcal{J}}

\newcommand{\bea}{\begin{eqnarray}}
\newcommand{\eea}{\end{eqnarray}}
\def\ol#1{\overline{#1}}

\documentclass{ws-ijgmmp}
\usepackage{tikz}
\usetikzlibrary{cd,patterns}
\usepackage[T1]{fontenc}
\newcounter{saveenumi}
\newcommand{\seti}{\setcounter{saveenumi}{\theenumi}}
\newcommand{\conti}{\setcounter{enumi}{\thesaveenumi}}
\newcounter{saveenumii}

\usepackage{cancel}

\begin{document}

\markboth{Authors' Names}
{Instructions for Typing Manuscripts (Paper's Title)}

%
\catchline{}{}{}{}{}
%

\title{Elementary methods for splitting representations of Rook monoids: a gentle introduction to groupoids.}

\author{Joseph Ben Geloun,$^{a,b,}$\footnote{bengeloun@lipn.univ-paris13.fr}\; \;     
Gérard H. E. Duchamp,$^{a,}$\footnote{gheduchamp@gmail.com}  \;\;  
Christophe Tollu$^{a,}$\footnote{ct@lipn.univ-paris13.fr}}

\address{$^a$Laboratoire d'Informatique de Paris Nord, UMR CNRS 7030\\
Univ. Sorbonne Paris Nord, 99 ave. J.-B. Clément, 93430, Villetaneuse, France\\
$^b$International Chair in Mathematical Physics and Applications (ICMPA-UNESCO Chair), 
072BP50, Cotonou, Benin
}
\maketitle
\begin{history}
\received{(Day Month Year)}
\revised{(Day Month Year)}
\end{history}

\begin{abstract}
We show that the algebra of the coloured rook monoid $R_n^{(r)}$, {\em i.e.} the monoid of $n \times n$ matrices with at most one non-zero entry (an $r$-th root of unity) in each column and row, is the algebra of a finite groupoid, thus is endowed with a $C^*$-algebra structure.  
This new perspective uncovers the representation theory of these monoid algebras by making manifest their decomposition in irreducible modules. 
\end{abstract}


\tableofcontents

\section{Introduction}
The importance of $C^*$-algebras for physics in general and quantum groups in particular cannot be overemphasized (see e.g. \cite{TT1,AVD}). In this paper, we offer yet another example of their usefulness in developing a general approach to the combinatorial representation theory of finite groupoids, which we present in detail below. On our way, we state and prove a general theorem saying that a minimal idempotent within the complex algebra of an isotropy group is also minimal in the algebra of the larger group. When combined with the Wedderburn-Artin formula, the latter result provides us with a procedure to construct a complete orthogonal family of minimal idempotents of the larger algebra.

We will use particular monoids of matrices known as rook matrices and a slight generalization thereof, which we call coloured rook monoids, as a benchmark to illustrate the efficiency of our approach. As one of its consequences, the combinatorial representation theory of coloured rook 
matrices boils down to the one of wreath products of cyclic groups with symmetric groups ($\mathbb{Z}_{r}\wr S_{k}$). The straightforwardness of our methodology and its combinatorial and even algorithmic stance enables us to avoid the use of the algebraic machinery often used in classical papers on the representation theory of the rook monoids and their deformations \cite{LS1, TH1, RP1}.

Rook monoids can be introduced in various ways (linear, set-theoretical or combinatorial, at least). The most simple one is to present them is as monoids of 
square $\{0,1\}$-matrices (of a given order) with at most one non-zero entry per row and column or, equivalently, permutation matrices in which a number of $1$'s have been replaced by $0$'s (see \cite{LS1} and \cite{GoF41,GoF100} for combinatorial aspects of the ‘‘rook rules''). These monoids, their structure and their representations have been extensively studied: see, for example, the very recent and lucid account by Mark Lawson, summing up a significant part of this enormous literature \cite{Lawson1}. Contrariwise, their coloured versions (resulting in a smash product of their algebras, see \cite{BoMa}) have not, to the best of our knowledge, been considered. However, the use of idempotents and Möbius function to exhibit the underlying $C^*$-algebra structure can still be carried over to the coloured case.

\section{The rook monoids $R_n$ and coloured rook monoids $R_n^{(r)}$.}\label{ColorRookMon}
This section presents the coloured rook monoids $R_n^{(r)}$, their specializations $R_n=R_n^{(1)}$ and their algebras $\C[R_n^{(r)}]$. 
We identify partial identities that are key elements revealing the underlying groupoid structure of $\C[R_n^{(r)}]$.
\subsection{Definitions}\label{RookDef}
For $n\in \N_{\geq 1}$, let $R_n\subset \Z^{n\times n}$ 
(or, more precisely, $\N^{n\times n}$, see \cite{Eil} Ch VI \S 2) 
be the set of 
$\{0,1\}$-matrices $M$ of size $n\times n$ having at most one non-zero entry per row and column. Likewise, for $r>0$, $R_n^{(r)}$ stands for the set of matrices defined identically except that the non-zero entries now range over the set of $r$-th roots of unity. For example, consider the four $3\times 3$ matrices defined by the following equations:
\begin{equation}\label{ex1}
M_1=\begin{pmatrix}
0 & 0 & 1\\
0 & 0 & 0\\
0 & 1 & 0\\
\end{pmatrix}\ \ 
M_2=\begin{pmatrix}
0 & 0 & 0\\
1 & 0 & 0\\
0 & 1 & 0\\
\end{pmatrix}\ \ 
M'_1=\begin{pmatrix}
0 & 0 & i\\
0 & 0 & 0\\
0 & -1 & 0\\
\end{pmatrix}\ \ 
M'_2=\begin{pmatrix}
0 & 0 & 0\\
-i & 0 & 0\\
0 & 1 & 0\\
\end{pmatrix}
\end{equation}
Clearly, $M_1, M_2\in R_3$ and $M'_1, M'_2\in R_3^{(4)}$. 
Each matrix in $R_n$ (every $\{0,1\}$-matrix, actually) is in 1-1 correspondence
with its {\it support}, the set of coordinates $(column, row)$ in $[1,\ldots ,n]\times [1,\ldots ,n]$ of the $1'$s. For example, the supports of the matrices of Eq. \eqref{ex1} are  
\begin{equation}\label{support1}
supp(M_1)=\big\{(2,3),(3,1)\big\}\ ;\ supp(M_2)=\big\{(1,2),(2,3)\big\}
\end{equation}
This support is always the graph of an injection $j:\ P\to Q$ with 
$P,Q\subset [1,\ldots ,n],\ |P|=|Q|=rk(M)$ ({\em i.e.} 
$dom(j)=s(M)=P,\ codom(j)=t(M)=Q$)\footnote{The sets $dom$ and 
$codom$ will soon become ‘‘source'' and ''target'', 
hence the notations $s(M),\,t(M)$.} and 
the $(column, row)$ coordinates are all of the form $(x,j(x))$. For example, the (partial) injections corresponding to matrices of Eq. 
\eqref{ex1} are\footnote{Each time $j_i$ corresponds to $M_i$.}
\begin{equation*}
dom(j_1)=\{2,3\},\ j_1(2)=3,\ j_1(3)=1\ ;\ 
dom(j_2)=\{1,2\},\ j_2(1)=2,\ j_2(2)=3
\end{equation*}
Supports for coloured matrices are defined likewise but the correspondence from $R_n^{(r)}$ to the  supports is many to one.

It is easily checked that $R_n^{(r)}$ is closed under matrix multiplication and contains the unit matrix $[I_{n\times n}]$, therefore it is a monoid of matrices. The monoid $R_n=R_n^{(1)}$ is nicknamed ‘‘rook monoid'' because the non-zero entries correspond to non-attacking rook positions on a chessboard. We will call $R_n^{(r)}$ the monoid of ‘‘coloured rook matrices'' because of the $r$-th roots of unity (see \cite{DuToPeKo} for the use of $r$-th roots of unity for colour factors, twisted products and $r$-ary supersymmetry).

One checks easily that the number of matrices of rank $k$ in $R_n^{(r)}$ is 
$\binom{n}{k}^2\,r^k.k!$ and, therefore, 
\begin{equation}\label{enum}
|R_n^{(r)}|=\sum_{k=0}^n\,\binom{n}{k}^2\,r^k.k!\ 
\end{equation}

Now, the elements of the $\C$-algebras of the monoids $R_n^{(r)}$ are formal linear combinations of elements of $R_n^{(r)}$ which are then {\it considered as linearly independant}. 
In other words, an element of $\C[R_n^{(r)}]$ is of the form $\sum_{A\in R_n^{(r)}}\,\alpha(A).A $, where $\alpha(A) \in \C$. In this respect, we warn the reader that matrices as elements of $\C[R_n^{(r)}]$ cannot be added as usual matrices, for example, in 
$\C[R_2^{(1)}]$, 
\begin{equation}\label{NoAdd}
\begin{bmatrix}
1 & 0\\ 0& 0
\end{bmatrix}
+ \begin{bmatrix}
0 & 0\\ 0& 1
\end{bmatrix}
\not= 
\begin{bmatrix}
1 & 0\\ 0& 1
\end{bmatrix}
\end{equation} 
that is why, in case of ambiguity, (coloured) rook matrices will be noted between square brackets, as in Eq. \eqref{NoAdd} above.
However, within $\C[R_n^{(r)}]$, matrices multiply ‘‘as matrices''
\begin{equation}\label{asMat}
\Big(\sum_{A\in R_n^{(r)}}\,\alpha(A).[A]\Big)\times 
\Big(\sum_{B\in R_n^{(r)}}\,\alpha(B).[B]\Big)=
\sum_{A,B\in R_n^{(r)}}\,\alpha(A)\alpha(B).[AB]
\end{equation} 
It is direct to check that the identity matrix $I_{n\times n}$ is the unity of 
$\C[R_n^{(r)}]$.
Note that, if $t(B)\not= s(A)$ (even if the ranks are equal), the rank of the product decreases. For this reason (and a more subtle reason of semi-simplicity which will be discussed later), we would like to keep only the principal part of this product, {\em i.e.} consider a new product $\ast$ defined by  
\begin{equation}
\label{StarProd1}
[A]\ast [B] = \left\{
\begin{array}{l} 
[AB] \;, \; \mbox{ if } \; t(B)= s(A) \\ 
  0 \;, \; \mbox{ otherwise.}  
\end{array}  \right. 
\end{equation}
The product $\ast$ yields a new $\C$-AAU (associative unital algebra) $(\C[R_n^{(r)}],\ast, \mathbf{1}_*)$, which turns out to be isomorphic to the old one, $(\C[R_n^{(r)}],\times,[I_{n\times n}])$ (in the sequel, the $\times$ product will be denoted by a simple dot ($\cdot)$). We expatiate on this isomorphism in Proposition \ref{TriangDef} below. Morover, there is a continuous deformation between the aforementioned algebras, as we plan to show in a forthcoming paper. 
\subsection{Partial identities: decoupling $\C[R_n^{(r)}]$}
\label{PartId}
Among the coloured rook matrices of size $n$, the so-called partial identities play the key rôle of local units. Indeed, for each $S\subset [1,\ldots ,n]$, let us denote 
$I_S=\sum_{i\in S}\,E_{ii}$ (here $E_{ii}\in R_n$ stands for the classical 
matrix units, {\em i.e.} the rook matrices such that $supp(E_{ii})=\{(i,i)\}$). Then $I_S$ is the rook matrix with incomplete diagonal and $supp(I_S)=\{(i,i)\}_{i\in S}$. For example, with $M_i$ as in Eq. \eqref{ex1}, one has 
\begin{equation}\label{InvArr}
^{t}\hspace{-1mm}M_1.M_1=I_{\{2,3\}},\ 
M_1.^{t}\hspace{-1mm}M_1=I_{\{1,3\}},\
M_2.^{t}\hspace{-1mm}M_2=I_{\{1,2\}},\
M_1.^{t}\hspace{-1mm}M_1=I_{\{2,3\}}
\end{equation} 
and, for a general $M\in R_n^{(r)}$, 
\begin{equation}\label{InvMon}
M^{\dag}.M=I_{s(M)},\ M.M^{\dag}=I_{t(M)}\mbox{ where }
M^{\dag}=^{t}\hspace{-1mm}\ol{M}
\end{equation}
Remark that the law of composition of these matrices is that of intersection of their indices, {\em i.e.} $I_X.I_Y=I_{X\cap Y}$. In fact, their set 
$\big\{I_S\big\}_{S\subset [1,\ldots ,n]}$ is a submonoid (call it $D_n\subset R_n$) isomorphic to the Boolean meet-lattice of subsets of $[1,\ldots ,n]$.\\   
Therefore, using the Möbius function, for every 
$A\subset [1,\ldots ,n]$, we set
\begin{equation}\label{E1}
e_A=\sum_{Q\subset A}(-1)^{|A\setminus Q|} [I_Q]
\end{equation} 
Note that this procedure (Eq. \eqref{E1}), very much in the style of the Murnaghan-Nakayama rule, can be extended to any coloured rook matrix $M$. So we set (the symbol $\subset$ denotes the inclusion between positions and values of non-zero entries),  
\begin{equation}\label{Rook-Möbius}
\mu(M):=\sum_{N\subset M}(-1)^{|M\setminus N|}\,[N]
\end{equation}
Using the exchange property, i.e. $I_{t(M)}M=MI_{s(M)}$, we can state 
\begin{proposition}\label{TriangDef}
We have the following\\ 
i) $\{e_A\}_{A\subset [1,\ldots,n]}$ is a complete set of orthogonal 
idempotents of 
$\Z[D_n]$, {\em i.e.} $\sum_{A\subset[1,\ldots,n]} = 1_{\mathbb{Z}[D_n]}$ and $e_A e_B =\delta_{A,B} e_A$.\\
ii) For all $M\in R_n^{(r)}$,
\begin{equation}
\mu(M):=M.e_{s(M)}=e_{t(M)}.M
\end{equation}
iii) For $M,N\in R_n^{(r)}$, one has
\begin{equation}\label{Separation}
\mu(M).\mu(N)=\mu(MN)\mbox{ if }t(N)=s(M)\mbox{ and }=0 \mbox{ otherwise }  
\end{equation}
(iv) $\mathbf{1}_*:= \mu^{-1}([I_{n\times n}]) = \sum_{S\subset [1,\ldots, n]} I_S$ is the unit for the product $\ast$ of Eq. \eqref{StarProd1}
\end{proposition}
We see, from Eq. \eqref{Separation}, that the product $\ast$ defined by Eq. \eqref{StarProd1} is the image of the dot ($\cdot$) product by the unitriangular transformation
\begin{equation}
[M]\to \mu(M)=M.e_{s(M)}=e_{t(M)}.M
\end{equation} 
(this transformation is unitriangular w.r.t. the basis $\{M\}_{M\in R_n^{(r)}}$, with any order compatible with ranks because $\mu(M)=[M]+\sum\,\mbox{\it lower ranks }$). 
\section{Generalities on groupoids}\label{GenGr}
A groupoid $\calG$ is a category\footnote{Indeed, here, an oriented (labelled) graph as this groupoid will be finite.} where each arrow admits an inverse. This will be made more precise with the following notations and data. We need 
\begin{enumerate}
\item A set of objects $Ob(\calG)$ and a set of arrows $Ar(\calG)$
\item Two maps\footnote{Precisely, ‘‘source'' and ‘‘target'' of the arrow.} $s,t:\ Ar(\calG)\TArrow{1cm}Ob(\calG)$ such that arrows $\alpha,\beta\in Ar(\calG)$ are composable iff $t(\alpha)=s(\beta)$. Their composition will be noted $\beta\circ\alpha$ or, for short, $\beta\alpha$.    
\item A map $x\mapsto 1_x:\ Ob(\calG)\to Ar(\calG)$ s.t. 
\begin{enumerate}
\item for all $x\in Ob(\calG)$, $s(1_x)=t(1_x)=x$; 
\item for all $\alpha\in Ar(\calG)$, 
$1_{t(\alpha)}\alpha=\alpha1_{s(\alpha)}=\alpha$. 
\end{enumerate}
\item For all $\alpha\in Ar(\calG)$, there exists (a unique) $\alpha^*\in Ar(\calG)$ s.t. 
$\alpha^*\alpha=1_{s(\alpha)}$ and $\alpha\alpha^*=1_{t(\alpha)}$.
\end{enumerate}
Eq. \eqref{StarProd1} and the results of subsection \ref{PartId} suggest that the coloured rook monoid $R_n^{(r)}$ can be arranged naturally into a groupoid. 
To this end, we partition $R_n^{(r)}$ according to the rank, so 
$R_n^{(r)}=\sqcup_{0\leq k\leq n}R_n^{(r,k)}$.
 
First consider the category itself
\begin{enumerate}
\item The set of arrows $Ar(R_n^{(r)})$ is $R_n^{(r)}$ 
and the set of objects is $Ob(R_n^{(r)})=\sqcup_{0\leq k\leq n}Ob(R_n^{(r,k)})$ where 
$Ob(R_n^{(r,k)})=\binom{[1,\ldots,n]}{k}$, {\em i.e.} the set of $k$-subsets of $[1,\ldots,n]$    
\item The two maps $s,t$ are the ones defined in subsection \ref{RookDef}, respectively $dom$ and $codom$. 
\item In view of the ‘‘desired product'' of Eq. \eqref{StarProd1}, we see that, if two arrows are not composable, their product is zero and, if they are, their product is exactly as in the coloured rook monoid.        
\item The local units are exactly the rook matrices $1_X$ for $X\subset [1,\ldots,n]$, where $1_X$ is the $\{0,1\}$ diagonal matrix with $1$'s in the positions $(i,i)$ for $i\in X$ and zeroes elsewhere. In terms of supports, $1_X$ is the matrix $M$ such that $supp(M)=\{(i,i)\}_{i\in X}$ (see paragraph before Eq. \eqref{support1}).   
\seti
\end{enumerate}
The inverse property (for examples, see Eq. \eqref{InvArr}) is given by Eq.  \eqref{InvMon}
\begin{enumerate}
\conti
\item For all $M\in R_n^{(r)}$, one has $M^{\dag}.M=I_{s(M)},\ M.M^{\dag}=I_{t(M)}$
\end{enumerate} 
In this way, the algebra $\C[R_n^{(r)}]$ is exactly the algebra of a groupoid as we will see in the coming section. 
\section{The $C^*$-algebra of a finite groupoid and the rook monoid algebra}\label{C-star}
Let now $\calG$ be a finite groupoid. The $\C$-algebra of $\calG$ (noted $\C[\calG]$) has domain the vector space $\C^{Ar(\calG)}$, {\em i.e.} the space of 
$\C$-formal linear combinations of arrows. The (bilinear) multiplication of two arrows $\alpha,\beta\in Ar(\calG)$ is defined by 
\begin{equation}
\alpha\beta\mbox{ if } \alpha,\beta\mbox{ are composable and } 0 
\mbox{ otherwise.}
\end{equation} 
The involution, usually referred to as the star operation, is given by $M^*:=^{t}\hspace{-1mm}\overline{M}$ and extended as follows: 
\begin{equation}
\Big(\sum_{\alpha\in Ar(\calG)}c_{\alpha}\alpha\Big)^*=
\sum_{\alpha\in Ar(\calG)}\ol{c_{\alpha}}\alpha^*
\end{equation}
It is quicker to introduce right now the trace operator $Tr$ of $\C[\calG]$ (which extends the usual trace of matrices) by 
\begin{equation}\label{Trace}
Tr\Big(\sum_{\alpha\in Ar(\calG)}c_{\alpha}\alpha\Big)=
\sum_{x\in Ob(\calG)}c_{1_x}
\end{equation}  
One immediately checks that, 
\begin{enumerate}
\item for $P=\sum_{\alpha\in Ar(\calG)}c_{\alpha}\alpha$, one has 
$Tr(P^*P)=\sum_{\alpha\in Ar(\calG)}|c_{\alpha}|^2$
such that $||P||=\sqrt{Tr(P^*P)}$ is the classical $\ell^2$ norm of the vector space $\C[\calG]$ with orthonormal basis $Ar(\calG)$.
\item 
$M\to \dfrac{1}{|Ob(\calG)|}Tr(M^*M)$ is a faithful state of the 
$^*$-algebra $\C[\calG]$
\end{enumerate}
Thus $\C[\calG]$ is endowed with a $C^*$-algebra structure, through the left regular representation.

\smallskip
We can conclude that, with the new product $\ast$ defined by 
Eq. \eqref{StarProd1}), the algebra $(\C[R_n^{(r)}],\ast,\mu^{-1}(I_{n\times n}))$ can be considered as the algebra of a groupoid, Therefore, it is a finite-dimensional $C^*$-algebra and, as such, it can be decomposed as a direct sum of matrix algebras (Wedderburn-Artin decomposition).   
\subsection{Wedderburn-Artin decomposition of the algebra of a finite groupoid $\C[G]$ and of $\C[R_n^{(r)}]$}
In this paragraph, we suppose that $\calG=G$ is a finite connected groupoid. For 
$x\in Ob(G)$, we will note $G_x$ the subgroupoid of arrows $\alpha$ such that 
$s(\alpha)=t(\alpha)=x$. $G_x$ is a group, called the isotropy group of $G$ at $x$. Using $x$ as an origin, we will also use a choice of representative arrows, {\em i.e.} an injective map $y\mapsto \alpha_y$ from $Ob(G)$ to $Ar(G)$ such that 
for all $y \in Ob(G)$, $\alpha_y\in Hom_G(x,y)$. Let us set   
$\calF_x=\{\alpha_y\}_{y\in Ob(G)}$. We also choose a CSOMI (complete set of orthogonal minimal idempotents\footnote{{\em i.e.} a set $\{e^{(1)},\ldots, e^{(m)}\}$ of idempotents such that 
\begin{equation}
\label{idemUnit}
\sum_{1\leq i\leq m}\,e^{(i)}=1\; \mbox{ and } \; e^{(i)}e^{(j)}=
\delta_{ij}\,e^{(i)} \;(\mbox{for all }  1\leq i,j \leq m)
\end{equation}
and which is maximal, among such sets, for the order of refinement.}
) 
$(e_x^{(i)})_{1\leq i\leq m}$ in $\C[G_x]$. 
The following theorem shows how to construct a 
complete orthogonal family of minimal idempotents
in $\C[G]$.
\begin{theorem}\label{WA-Dec}
Let $G$ be a finite connected groupoid. We have 
\begin{enumerate}
\item For all $x\in Ob(G)$, $1_x\C[G]1_x=\C[G_x]$ and the sum 
$\calA=\oplus_{x\in Ob(G)}\,1_x\C[G]1_x$ is a sub-$\C$-AAU of $\C[G]$ (referred to as the diagonal subalgebra of $\C[G]$). 
\label{WA1}
\item If $e\in \calA$ is a minimal idempotent, then there exists a unique $x\in Ob(G)$ such that $e\in 1_x\C[G]1_x=\C[G_x]$. 
\label{WA2}
\item For $e\in \C[G_x]$ as above, the left ideal $\C[G].e$ is 
equal to 
\begin{equation}\label{LeftMinId}
\C[G].e=(\C.\calF_x).\C[G_x].e \simeq \C.\calF_x)\otimes_{\C} \C[G_x].e
\end{equation} 
\label{WA3}
\item $\C[G].e\simeq (\C.\calF_x)\otimes_{\C} \C[G_x].e$
\label{WA4}
\item For each $x\in Ob(G)$, let $\{e_x^{(i)}\}_{1\leq i\leq m}$ be a complete orthogonal family of minimal idempotents of $\C[G_x]$. We have the following decompositions in left ideals
\begin{equation}
\C[G].1_x=\oplus_{1\leq i\leq m}\, \C[G].e_x^{(i)}\mbox{ and }
\C[G]=\oplus_{y\in Ob(G)}\C[G].1_y
\end{equation}
\label{WA5}
\item $\C[G]=\oplus_{y\in Ob(G)}\C[G].1_y$
\label{WA6}
\item We can get a CSOMI 
of $\C[G]$ as follows: 
\begin{enumerate}
\item Choose a particular $x\in Ob(G)$ and $\{e^{(i)}\}_{1\leq i\leq m}$ a complete set of minimal orthogonal idempotents of $\C[G_x]$
\item For all $y\in Ob(G)$, set $e_y^{(i)}:=\alpha_y e_x^{(i)}\alpha_y^*$. \label{WA7b}
\end{enumerate}
Then
\begin{enumerate} 
\setcounter{enumii}{2}
\item The family $(e_y^{(i)})_{1\leq i\leq m}$ is a complete set of (orthogonal) minimal idempotents in $\C[G_y]$.
\label{WA7c}
\item The set $\{e_y^{(i)}\}_{1\leq i\leq m,y\in Ob(G)}$ is a complete set of (orthogonal) minimal idempotents in $\C[G]$. 
\label{WA7d}
\end{enumerate}
\end{enumerate} 
\end{theorem}
\proof 
\eqref{WA1} The only thing to check is $\sum_{x\in Ob(G)}\,1_x=1_{\C[G]}$, 
which is direct. \\
\eqref{WA2} The idempotents $\{1_x\}_{x\in Ob(G)}$ being central in $\calA$, then every idempotent $e\in \calA$ splits as a sum (of idempotents) 
$e=\sum_{x\in Ob(G)}\,e.1_x$.\\ 
\eqref{WA3} If $\beta\in \C[G]$, with $s(\beta)=x$ and $t(\beta)=y$, then the identity $\beta.e=\alpha_y(\alpha_y^*\beta).e$, with $\alpha_y \in \calF_x$, holds, which is the expected decomposition.\\
\eqref{WA4} This is due to the fact that 
$\C.\calF_x=\oplus_{y\in Ob(G)} \C.\alpha_y$ \\
\eqref{WA5} This is an immediate consequence of \eqref{WA3} and \eqref{WA4}.\\   
\eqref{WA6} The property falls from $1_{\C[G]}=\oplus_{y\in Ob(G)}\,1_y$.\\   
\eqref{WA7c} The property is obtained by a direct computation.\\   
\eqref{WA7d} The family of minimal idempotents $\{e_y^{(i)}\}_{1\leq i\leq m,y\in Ob(G)}$ is complete and orthogonal because of $1_{\C[G]}=\oplus_{y\in Ob(G)}\,1_y$, (\ref{WA7c}) and the orthogonality of the $1_y$.\\ 
Let $F$ be the set of minimal central idempotents of $\C[G]$, we have 
\begin{equation}
1_{\C[G]}=\sum_{x\in Ob(G)}\,1_x=\sum_{f\in F}\,f
\end{equation}
note that $F$ is in 1-1 correspondence with the types of minimal modules of 
$\C[G]$. In order to establish that the idempotents $e_y^{(i)}$ are minimal, we first need to know that each of the left ideals $\C[G].e_y^{(i)}$ is isotypic (\cite{Bou-Alg-8} \S 4.4).\\ 
Thus, let $e_y^{(i)}$ be one of the idempotents in the list of \eqref{WA7d}. 
We have to preserve the order of refinement and, in particular, avoid ‘‘outer'' decompositions of the type $e=(e_1+\alpha)+(e_2-\alpha)$ where $e=e_1+e_2$ and $\alpha\in Hom(y,z)$ with $z\not=y$. To this end, we use central idempotents.\\
It is clear that, because of 
\begin{equation}
e_y^{(i)}=e_y^{(i)}.1_{\C[G]}=\sum_{f\in F}\,e_y^{(i)}.f
\end{equation} 
there exists one $f\in F$ such that $e_y^{(i)}.f\not= 0$. 
Then, we write 
$e_y^{(i)}=e_y^{(i)}.f+e_y^{(i)}.(1-f)$ and notice that, due to the centrality of $f$, we have  
$$
1_y\,e_y^{(i)}.f\,1.y=1_y\,e_y^{(i)}\,1.y.f=e_y^{(i)}.f\in \C[G_y]
$$
The same computation holds for $e_y^{(i)}.(1-f)$, which proves that the decomposition 
$e_y^{(i)}=e_y^{(i)}.f+e_y^{(i)}.(1-f)$ occurs within $\C[G_y]$. We can then apply the fact that $e_y^{(i)}$ is minimal within $\calA$ to get $e_y^{(i)}=e_y^{(i)}.f$. 
This proves that $\C[G].e_y^{(i)}=\C[G].e_y^{(i)}.f$ is isotypic (\cite{Bou-Alg-8} \S 4.4).\\
The minimality of the idempotents is a consequence of the following lemma, which we state without proof.
\begin{lemma}\label{Isotype}
Let $\calA$ be a semisimple algebra over an algebraically closed field. Let $T$ be the set of types of $\calA$-modules (isomorphism classes). We consider a decomposition 
\begin{equation}\label{Adec}
\calA=\oplus_{j\in I}\,\calJ_j
\end{equation}
into isotypic (left) ideals and the equivalence 
\begin{equation}
j_1\sim_I j_2\Longleftrightarrow \calJ_{j_1}\sim_{\calA\mbox{-mod}}\calJ_{j_2}
\end{equation}
Let $J$ be a section of $\sim_I$.
Then\\
i) $\dim(\calA)\leq\,\sum_{j\in J}\,\dim(\calJ_j)^2$.\\
ii) if these two quantities are equal, then all $\calJ_{j}$ are minimal left ideals. \hfill$\square$
\end{lemma}
To complete the proof, let us remark that $\dim(\C[G])=|Ar(G)|=|Ob(G)|^2.|G_x|$ on the other hand, due to the fact that the family $\big(e_y^{(i)}\big)_{1\leq i\leq m,y\in Ob(G)}$ is a complete, we have 
\begin{equation}
\C[G]=\oplus_{1\leq i\leq m\atop y\in Ob(G)}\,\C[G].e_y^{(i)} 
\end{equation} 
but Th. \ref{WA-Dec}.\eqref{WA4} implies that $\dim(\C[G].e_y^{(i)})=|Ob(G)|.\dim(\C[G_y].e_y^{(i)}$). 
We conclude the proof of minimality by calling Lemma \ref{Isotype}. 
\hfill$\square$

\medskip
\noindent
Let us now indicate how to apply Theorem \ref{WA-Dec} 
to the decomposition of 
$\C[R_n^{(r)}]$. 
\begin{itemize}
\item Firstly, transform the monoid algebra $\C[R_n^{(r)}]$ into the algebra of a groupoid through the isomorphism of Eq. \eqref{StarProd1}
\item The new algebra $(\C[R_n^{(r)}],\ast,\mu^{-1}([I_{n\times n}]))$ is the direct sum of the algebras of the connected components w.r.t. the partition of $R_n^{(r)}$ according to the rank. Explicitly, one has $R_n^{(r)}=\sqcup_{0\leq k\leq n}\,R_n^{(r,k)}$.   
\item For a given rank $k\leq n$, $\C[R_n^{(r,k)}]$ is the algebra of a connected groupoid with objects $\binom{[1,\ldots ,n]}{k}$, {\em i.e.} the set of $k$-subsets of $[1,\ldots ,n]$, and hom-sets $Hom(X,Y)$ the sets of bijections $X\to Y$. 
\item For $k\leq n$ as above, we choose $X_k=[1,\ldots ,k]$ as origin and 
$\calF_{X_k}=\{\alpha_Y\}_{|Y|=k}$ where $\alpha_Y$ is the unique increasing bijection $[1,\ldots ,k]\to Y$.
\item To complete the setting, choose a CSOMI (complete set of orthogonal minimal idempotents) in $\C[G_{X_k}]\cong \C[\mathbb{Z}_{r}\wr S_{k}]$ and its elements with $\alpha_Y$ and $\alpha_Y^*$ as in Th. \ref{WA-Dec}.\eqref{WA7b}.  
\end{itemize}
\section{Concluding remarks}
Theorem \ref{WA-Dec} shows that minimal idempotents of $\C[R_n^{(r)}]$ can be found within the minimal idempotents of $\C[G_x]$ for all $x\in Ob(G)$: pick one in each connected component and then conjugate. The groups $G_x$ are the wreath products 
$\mathbb{Z}_{r}\wr S_{k}$ for all ranks $0\leq k\leq n$. The character theory of these wreath products was recently clarified by Adin and Roichman who gave a Murnaghan-Nakayama rule for them \cite{Wreath2}. Eq. \eqref{InvMon} indicates that $R_n^{(r)}$ in an inverse monoid. More generally, one can replace the roots of unity by the elements of an arbitrary finite group $G$ (with complex conjugates $g^*$ replaced by inverses $g^{-1}$) and still turn $R_n^{(G)}$ into a finite inverse monoid whose representation theory boils down to that of the wreath products $G\wr S_{k}$.

General results like that of Penrose and Munn \cite{Munn1} about the class of ‘‘inverse monoids'' opened a wide area of research (see the recent account by Lawson \cite{Lawson1} and the relationships between early semigroup theory \cite{CP1} and category theory showing up in \cite{Steinberg1}).

In closing, let us mention that from the facts that the idempotents $e_A$ belong to 
$\Z[R_n^{(G)}]$ and that $\kk[R_n^{(G)}]=\kk\otimes_\Z\,\Z[R_n^{(G)}]$, one infers 
that for every field $\kk$ of characteristic zero, the algebra 
$\kk[R_n^{(G)}]$ is semisimple, which shows that our study can be extended to other fields of coefficients.


\begin{thebibliography}{0}
%
\bibitem{Bou-Alg-8} Bourbaki, {\it Algebra, Chapter 8}, Springer, 2023 (1st ed. 2022). 
%
\bibitem{BoMa} A. Borowiec and W. Marcinek, \textit{On crossed product of algebras,} J. Math. Phys. \textbf{41} (2000) 6959-6975.
%
\bibitem{TT1} T. Timmermann, {\it An invitation to quantum groups and duality: from Hopf algebras to multiplicative unitaries and beyond}, European Mathematical Society, 2008
%
\bibitem{AVD} A. Van Daele, {\it From Hopf algebras to topological quantum groups: A short history, various aspects and some problems}, 
[\texttt{arXiv:1901.04328}]
%
\bibitem{LS1} L. Solomon, {\it Representations of the rook monoid}, J. Algebra \textbf{256} (2002) 309–342. 
%
\bibitem{TH1}T. Halverson, {\it Representations of the $q$-rook monoid}, J. Algebra {\bf 273} (2004) 227–251.
%
\bibitem{RP1}R. Paget, {\it Representation theory of $q$-rook monoid algebras}, J. Algebr. Comb. {\bf 24} (2006) 239–252.
%
\bibitem{GoF41} P. Blasiak, A. Horzela, G. H. E. Duchamp, K. A. Penson, and A. I. Solomon, {\it Heisenberg-Weyl algebra revisited: Combinatorics of Words and Paths}, J. Phys. A: Math. Theor. \textbf{41} (2009) 415204. [\texttt{arXiv:0904.1506}] 
%
\bibitem{GoF100} A.I. Solomon, P. Blasiak, G. Duchamp, A. Horzela and K.A. Penson, {\it Normal Order: Combinatorial Graphs}, 
Proc. 3rd Int. Symp. Progress in Supersymmetric Quantum Mechanics, 
World Scientific Publishing, 2004. [\texttt{arXiv:quant-ph/0402082}]
%
\bibitem{Lawson1} M. V. Lawson, {\it Introduction to inverse semigroups}, 2023. [\texttt{arXiv:2304.13580}] 
%
\bibitem{Eil} S. Eilenberg, \textit{Automata, Languages, and Machines,  Vol. A}, Academic Press, 1974.
%
\bibitem{DuToPeKo} G. H. E. Duchamp, C. Tollu, K. A. Penson and G. A. Koshevoy, {\it Deformations of Algebras: Twisting and Perturbations}, Séminaire Lotharingien de Combinatoire, {\bf B62e} (2010).
%
\bibitem{Wreath2} R. M. Adin and Y. Roichman, {\it On characters of wreath products}, Combinatorial Theory {\bf 2} (2) (2022), \#17. [\texttt{arXiv:2107.11899}]
%
\bibitem{Munn1} W. D. Munn and R. Penrose, {\it A note on inverse semigroups}, Math. Proc. Camb. Philos. Soc. {\bf 51} (1955) 396–-399. 
%
\bibitem{CP1} A. H. Clifford and G. B. Preston, {\it The algebraic theory of semigroups, vol. 1.}, Mathematical Surveys and Monographs, No. 7, Amer. Math. Soc., 1961.
%
\bibitem{Steinberg1} A. Costa and B. Steinberg, 
{\it The Schützenberger category of a semigroup}, 
Semigroup Forum {\bf 91} (2015), 543–559. [\texttt{arXiv:1408.1615}]
%
\bibitem{JK1} G. D. James and A. Kerber, {\it The representation theory of the symmetric group}, Encycl. of Math. vol. 16, Addison-Wesley, 1981.
%
\bibitem{MO184509} Isotypic components of the action of the symmetric group on polynomials. [\texttt{https://mathoverflow.net/questions/184509}]
%
\end{thebibliography}
\end{document}